\documentclass[fleqn,10pt]{wlscirep}
\usepackage[utf8]{inputenc}
\usepackage[T1]{fontenc}
\title{Examination of flow birefringence induced by the shear components along the optical axis using a parallel-plate-type rheometer}

\author[1]{William Kai Alexander Worby}
\author[1]{Kento Nakamine}
\author[2]{Yuto Yokoyama}
\author[3*]{Masakazu Muto}
\author[1**]{Yoshiyuki Tagawa}
\affil[1]{Dept. of Mechanical Systems Engineering, Tokyo University of Agriculture and Technology, Koganei, Tokyo, 184-8588, Japan}

\affil[2]{Micro/Bio/Nanofluidics Unit, Okinawa Institute of Science and Technology Graduate University, Kunigami-gun, Okinawa, 904-0495, Japan}

\affil[3]{Dept. of Electrical and Mechanical Engineering, Nagoya Institute of Technology, Nagoya, Aichi, 466-8555, Japan}

\affil[*]{\ E-mail:muto.masakazu@nitech.ac.jp}
\affil[**]{E-mail:tagawayo@cc.tuat.ac.jp (Corresponding author)}

\keywords{Flow birefringence, Rheo-optics, Photoelasticity}

\begin{abstract}
In this study, the concept of rheo-optics is applied that explores the flow birefringence caused by stress components along the optical axis of the camera since it is often overlooked in the traditional theories of photoelastic flow measurement. A novel aspect of this research is that it involved conducting polarization measurements on simple shear flows, specifically from a perspective in which a shear-velocity gradient exists along the camera's optical axis. A parallel-plate-type rheometer and a polarization camera are employed for these systematic measurements. The experimental findings for dilute aqueous cellulose nanocrystal suspensions demonstrates that the flow birefringence can be expressed as a power law based on the power of the second invariant of the deformation-rate tensor. This suggests that flow birefringence can be universally characterized by the coordinate-independent invariants and a pre-factor determined by the direction of polarization measurement. By adjusting the nonlinear term in the stress-optic law, its applicability could be expanded to include three-dimensional fluid stress fields in which the stress is distributed along the camera's optical axis.
\end{abstract}

\usepackage{upgreek}

\begin{document}
\flushbottom
\maketitle

\newcommand{\B}[1]{{\textcolor{blue}{#1}}}
%

\section*{Introduction}\label{sec:introduction}
The non-invasive measurement of the internal stress states of fluids in flow is of great importance for flow engineering and biomechanics.
For bulk fluid pressure measurements, the method of measuring velocity fields with particle tracking velocimetry\cite{Adamczyk1988} or particle image velocimetry\cite{Oudheusden2013} is commonly used.
However, they require spatial derivatives of velocity fields to determine the stress, in which the noise amplification problem appears.
Other methods have also been proposed to evaluate the information corresponding to the stress based on unique phenomena, such as light scattering and magnetic resonance\cite{McAfee1974, Martins1986, Odagiri2016}.
These methods can convey information relating to the stress driving the fluid intact.
Among these approaches, measurement of stress or velocity gradient distributions by the photoelastic method\cite{Aben2012} has been applied and is attractive due to its high measurement sensitivity and experimental simplicity.

The photoelastic method was developed as a stress-measurement technique for solids, and it has been further developed and widely studied for 50~years\cite{Lautre2015}. 
Among solid materials, various studies considering the measurement of stresses and residual stresses, especially in glass, have been undertaken\cite{Aben2012, Ramesh2021}. 
Stress-loaded materials change their refractive index with regard to the direction of polarization vibration in response to strain. 
Therefore, when two orthogonally polarized light beams are transmitted through a strained material, retardation will appear between them. 
The incident and emitted polarized light is represented by a composite vector of the two linearly polarized light beams, and their trajectories will differ due to the retardation of the composite vector. 
Depending on the trajectory difference, the result can be classified as linearly, elliptically, or circularly polarized light.
When circularly polarized light is incident onto a stress-loaded material, it will be emitted as elliptically polarized light with retardation $\Delta$ and orientation angle $\phi$. 
The values of $\Delta$ and $\phi$ (photoelastic parameters) correspond to the principal-stress difference and the principal-stress direction, respectively\cite{Aben2012}. 
In the photoelastic method, stress can be estimated from $\Delta$ based on the stress-optic law (SOL)\cite{Prabhakaran1975, Ramesh2021}:
\begin{equation}
    \Delta = \int C_1(\sigma_1 - \sigma_2)~{\rm d}h,
    \label{eq:SOL1}
\end{equation}
where: ${\rm d}h$ is the infinitesimal thickness of the material; $C_1$ is the stress-optic coefficient; and $\sigma_1$ and $\sigma_2$ are the maximum and minimum principal stresses, respectively. 
However, Eq.~(\ref{eq:SOL1}) holds only for two-dimensional (2D) stress fields with no stress along the optical axis, or, if there is stress along the optical axis, then it must be uniform.
This means that $\Delta$ can be obtained by only considering the secondary principal stress difference\cite{Doyle1978, Sampson1970}. 
Here, the secondary principal stress difference ($\sigma_1 - \sigma_2$) is the principal stress difference projected onto a plane perpendicular to the camera's optical axis. 
In the case of 3D stress fields, applying the SOL is more complicated than in the above equation. 
For stress distributions along the camera's optical axis, it is necessary to introduce the concepts of the ``optically equivalent model''\cite{Srinath1974} and ``integrated photoelasticity''\cite{Ramesh2021, Yokoyama2023}. 
This makes it possible to consider that the 3D stress fields consist of sufficiently thin linear polarizers that can be assumed to be 2D stress fields. 
Thus, the polarization state transmitting through the 3D stress field can be calculated by multiplying the Mueller matrices of each optically equivalent model\cite{Yokoyama2023}.

Recently, studies have been conducted on the application of photoelastic methods to fluids, and it has been shown the principle of Eq.~(\ref{eq:SOL1}) can be applied to quasi-2D flows\cite{Noto2020, Lane2023}. 
However, the possibility of applying photoelastic methods to 3D fluid stress fields has not yet been fully discussed. 
As the flow becomes more 3D, the stress distribution along the camera's optical axis (hereafter simply the ``optical axis'') becomes more significant. 
The effect of this stress distribution on the retardation is most obvious near the centre of the channel flow. 
Specifically, the retardation measurements obtained by integration along the optical axis deviate from zero.
These experimental findings have also been verified by numerical calculations, which established that the regions in which the impact of 3D effects are found to be dominant are close to the plane of symmetry\cite{Clemeur2004}. 
This phenomenon has been reported in circular and rectangular channels with aspect ratios close to 1\cite{Kim2017,Alizadehgiashi2018,Nakamine2024}. 
It has been concluded that this is due to shear on the upper and lower surfaces of the channel, although quantitative insights were limited.
Systematic investigations still need to identify the effect of the stress distribution along the optical axis on photoelastic parameters, intending to apply photoelastic methods to 3D fluid stress fields.

Investigations of the optical-anisotropy responses to the shear rate (shear stress) have been conducted using rheo-optics. 
Starting from the apparatus proposed by Lodge\cite{Lodge1956} and Philipoff\cite{Philippoff1961}, the filament stretching rheometer\cite{Rothstein2002} and the capillary breakup extensional rheometry dripping-onto-substrate (CaBER-DoS) method\cite{Muto2022,Muto2024} were developed to perform birefringence measurements on liquid polymers under uniaxial extension flow. 
Additionally, the use of rheometers in shear flow has also been reported in many studies; as optical-anisotropy measurement systems are easy to incorporate into experimental apparatus, there have been reports of the use of rheometers with different measurement principles, including the concentric cylinder (CC)-type\cite{Tanaka2018,Lane2022}, the parallel plate (PP)-type\cite{Kadar2020,Detert2023}, and the cone plate (CP)-type\cite{Oba2016}. 
The optical-anisotropy induced by secondary principal stress difference can be measured from the shear-vorticity direction, e.g., in uniaxial extension systems and CC-type rheometers. Conversely, measurements from the shear direction provide a direct visualization of the stress along the optical axis rather than showing the principal stress difference. 
This is due to the existence of a velocity gradient along the optical axis, which is of interest for the present study. 
Most previous rheo-optical measurement cases using PP-type rheometers have considered non-Newtonian fluids, such as molten polymers\cite{Mykhaylyk2016,Hausmann2018,Kadar2020}. 
In general, the flow behaviour of non-Newtonian fluids differs considerably from that of Newtonian fluids. 
Therefore, any measurements will include not only shear but also elastic and normal stress contributions. 
Evaluating birefringence after isolating these stresses is a complicated and challenging process.

In the present study, photoelastic measurements were performed in the direction of shear on a simple Couette flow using a PP-type rheometer. 
To simplify the discussion, we used a fluid that can be assumed to be a Newtonian fluid as the first step.
We used a dilute suspension showing Newtonian behaviour as the birefringent fluid and attempted to extract the stress components along the optical axis.
The tendency of birefringence induced by the stress along the optical axis was quantitatively examined and compared with the results obtained in previous studies. 

\section*{Principles}
This section outlines the basic theory of polarization measurements, mainly focusing on flow birefringence and the SOL\cite{Prabhakaran1975}.

\subsection*{Flow birefringence of fluids}\label{sec:FB}
\begin{figure}[tb!]
\begin{center}
\includegraphics[width=0.6\textwidth]{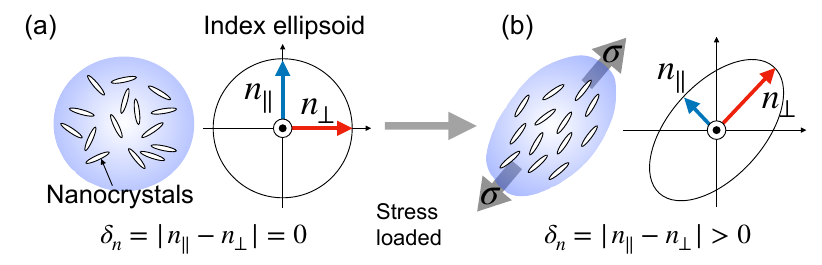}
\caption{Schematic of nanocrystals in solution (a)~when no stress is applied, (b)~under stress loading. Although needle-like nanocrystals are shown, the mechanism of onset is the same with macromolecular chains.}
\label{fig:FIB}
\end{center}
\end{figure}

Birefringent fluids are composed of crystals or polymer chains with large aspect ratios. This is the key to changing the refractive indices $n_{\perp}$ and $n_{\parallel}$ perpendicular and parallel to the direction of vibration of the transmitted light, respectively. One of the optical anisotropies, birefringence $\delta_n$, is a physical quantity that indicates the magnitude of the anisotropy of the refractive index. Birefringence is defined as the absolute value of the difference between the major and minor diameters of the index ellipsoid:
\begin{equation}
\delta_n = |n_{\perp} - n_{\parallel}|.
\label{eq:birefringence}
\end{equation}
As long as there is no stress loading, birefringent fluids show optically isotropic properties ($n_{\perp} = n_{\parallel}$) because the particles are randomly oriented by Brownian motion. However, when shear is applied, the crystals or polymers become aligned in the direction corresponding to the stress, resulting in optical-anisotropy ($n_{\perp} \neq n_{\parallel}$). When the particles are strongly oriented in a particular direction, the anisotropy of the refractive index becomes stronger and the value of the birefringence increases (see Fig.~\ref{fig:FIB}a). Conversely, when the applied shear is reduced, the orientation becomes random again and birefringence is no longer induced (see Fig.~\ref{fig:FIB}b). This phenomenon is known as flow birefringence\cite{Maxwell1874}; as examples of birefringent fluids, aqueous cellulose nanocrystal (CNC) suspensions\cite{Calabrese2021}, wormlike micelle solutions\cite{Ober2011}, and xanthan gum solutions\cite{Yevlampieva1999} are well known for showing birefringence.

\subsection*{Stress-optic law (SOL)}
The retardation $\Delta$ obtained by the photoelastic method is the summation (integrated value) of the birefringence along the optical axis. The retardation $\Delta$ caused by flow birefringence $\delta_n$ is related to the strain rate $\dot e_{ij}$ inside the fluid\cite{Doyle1982}:
\begin{equation}
\Delta{\rm cos}2\phi=\int \alpha_1({\dot e_{{\it yy}}}-{ \dot e_{{\it xx}}})+\alpha_2[(\dot e_{yy}+\dot e_{xx})(\dot e_{yy}-\dot e_{xx})+\dot e_{zy}^2-\dot e_{xz}^2]~{\mathrm d}z,
\label{eq:sol-A}
\end{equation}
\begin{equation}
\Delta\rm{sin}2\phi=\int 2\it\alpha_{\rm1}\dot e_{xy}+\alpha_{\rm 2}[\rm{2}\it(\dot e_{yy}+\dot e_{xx})\dot e_{xy}+\rm{2}\it \dot e_{xz}\dot e_{yz}]~{\mathrm d}z.
\label{eq:sol-B}
\end{equation}
Here, the optical axis is defined as the $z$ axis of a Cartesian coordinate system, and $\alpha_1$ and $\alpha_2$ are functions of the physical properties of the fluid. For Newtonian fluids, the stress is proportional to the strain rate. Therefore, Eqs.~(\ref{eq:sol-A}) and (\ref{eq:sol-B}) can be expressed using stress\cite{Nakamine2024}:
\begin{equation}
\Delta {\rm cos} 2\phi=\int {\it C}_1({\sigma_{{\it yy}}}-{\sigma_{{\it xx}}})+C_2[( \sigma_{yy}+ \sigma_{xx})(\sigma_{yy}- \sigma_{xx})+ \sigma_{zy}^2- \sigma_{xz}^2]~{\mathrm d}z,
\label{eq:sol-C}
\end{equation}
\begin{equation}
\Delta{\rm sin}2\phi=\int 2\it C_{\rm 1} \sigma_{xy}+C_{\rm 2}[\rm{2}\it( \sigma_{yy}+ \sigma_{xx}) \sigma_{xy}+\rm{2}\it \sigma_{xz} \sigma_{yz}]~{\mathrm d}z.
\label{eq:sol-D}
\end{equation}
In these equations, $C_1 =\alpha_1/\eta$ and $C_2 = \alpha_2/\eta^2$, in which $\eta$ is the shear viscosity of the fluid. Aben and Puro\cite{Aben1997} also discussed the optical relationship based on Eqs.~(\ref{eq:sol-C}) and (\ref{eq:sol-D}) and assumed that the stress components along the optical axis, i.e., $\sigma_{xz}$, $\sigma_{zy}$, and $\sigma_{yz}$, were negligible. In other words, they made the assumption that $C_2 = 0$, which leads to the proposal of:
\begin{equation}
\Delta = \int C_1\sqrt{(\sigma_{xx} - \sigma_{yy})^2 + 4{\sigma_{xy}}^2}~{\mathrm d}z.
    \label{eq:SOL}
\end{equation}
It should be emphasized again that Eq.~(\ref{eq:SOL}) is a relation that holds only for 2D stress fields. Moreover, Eq.~(\ref{eq:SOL}) is an often-used expression in the solid-state photoelastic method\cite{Riera1969,Aben1997}.

\section*{Methodology}
In this section, the details of the experiments are presented. Unless otherwise indicated, all experiments were repeated at least three times per case at 25$^\circ$C.

\subsection*{Experimental setup}
\begin{figure*}[tb!]
\begin{center}
\includegraphics[width=0.9\textwidth]{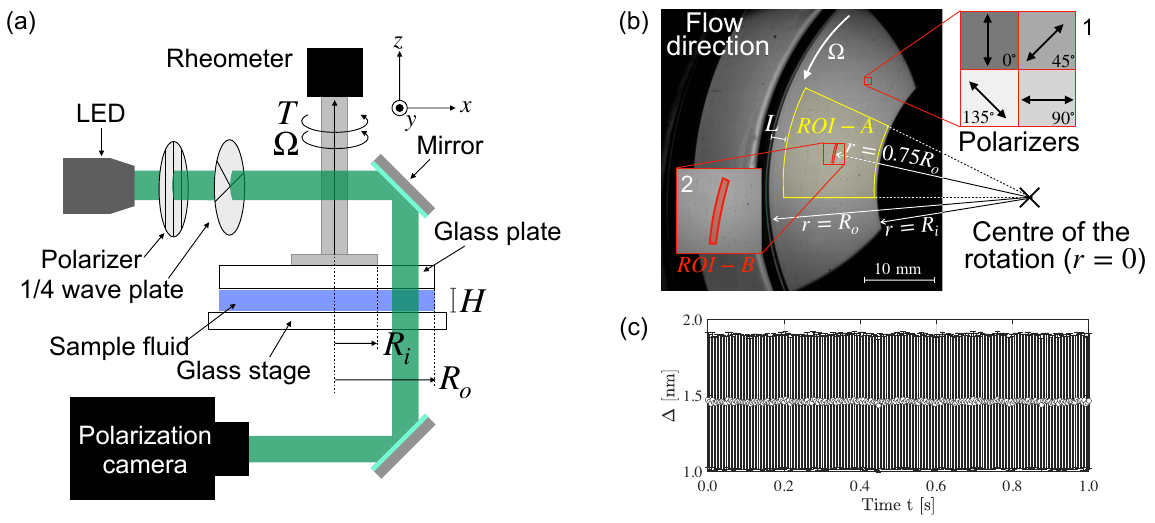}
\caption{(a)~Schematic of the rheo-optical setup, in which $R_i$ and $R_o$ are the inner and outer radii of the transparent part of the plate, respectively. (b)~Representative intensity image for CNC suspension 1.0~wt\% at shear rate $\dot\gamma = 1000$~s$^{-1}$. As shown in inset~1 of panel~(b), retardation is obtained from four neighbouring polarizers. (c)~Temporal evolution of retardation at no flow.}
\label{fig:Setup}
\end{center}
\end{figure*}

Figure~\ref{fig:Setup}a shows a schematic of the rheo-optical measurement system. The stress-controlled rheometer (MCR~302, Anton Paar Co., Ltd.) was equipped with a parallel plate (PP43/GL-HT, Anton Paar Co., Ltd., plate radius $R_o = 21.5$~mm) made of quartz glass and a flat glass plate (PTD200/GL, Anton Paar Co., Ltd.). In this system, shear flow is induced by clockwise rotation of the plate, and the average shear stress and torque are logged. Left-handed circularly polarized light is generated by attaching a polarizer and 1/4-wave plate to an LED light source (SOLIS-525C, Thorlabs Co., Ltd., wavelength $\lambda = 525$~nm). This polarized light is reflected from the top of the plate by a mirror and emitted as elliptically polarized light with retardation $\Delta$ and orientation angle $\phi$. The transmitted elliptically polarized light is reflected by the mirror again and enters the polarization camera (CRYSTA~PI-1P, Photron Co., Ltd.), which is equipped with a 524-nm band-pass filter. The polarization camera has a spatial resolution of up to $512 \times 512$~pixels and a temporal resolution of 1.55~Mfps. In these experiments, all measurements were made with a resolution of $512\times 512$~pixels (44.5~$\upmu$m/pixel) at 1000~fps. Additionally, the gap height between the plates was always fixed at $H = 100\ \pm 5$~$\upmu$m, and the shear rate was set to 1000--10,000~s$^{-1}$. Note that the shear rate is specified at 2/3	 of the plate radius. This is the range in which it is expected that no artefacts will appear in the rheological measurement results (for further details, please see the \ref{sec:window}).

\subsection*{Polarization camera}
\label{sec:CRYSTA}
A polarization camera (CRYSTA~PI-1P, Photron Co., Ltd.) was used to detect the retardation $\Delta$, which is the integrated value of birefringence along the optical axis of the light transmitted through the apparatus. Using the phase-shifting method\cite{Onuma2014}, this was obtained from the radiance through linear polarizers oriented in four different directions (0$^\circ$, 45$^\circ$, 90$^\circ$, and 135$^\circ$) in an area of $2 \times 2$~pixels (as shown in inset~1 of Fig.~\ref{fig:Setup}b). Defining the light intensities detected at each of these pixels as $I_1$, $I_2$, $I_3$, and $I_4$, respectively, the retardation can then be given by:
\begin{equation}
\Delta = \int\delta_n\ \mathrm{d}z=\frac{\lambda}{2\pi}{\rm sin^{-1}}\frac{2\sqrt{(I_3-I_1)^2+(I_2-I_4)^2}}{I_1+I_2+I_3+I_4},
     \label{phase-shift}
\end{equation}
where $\lambda$ [m] is the wavelength of the light source. As the retardation is obtained from four linear polarizers, the spatial resolution was $512 \times 512$~pixels, which is 1/4 of the $1024 \times 1024$-pixel light-intensity image. As there is no stress distribution along the optical axis, the retardation $\Delta$ can be calculated as the product of the birefringence $\delta_n$ and the gap height $H$. Therefore, in this study, we calculated the birefringence by dividing the retardation measurements by $H$.

\subsection*{Data acquisition and analysis}
An arc-shaped region of interest (ROI-A) located at a distance $L\approx 3$~mm from the edge of the plate (outlined in yellow in Fig.~\ref{fig:Setup}b) was chosen, and the background-subtracted retardation field during flow was provided by the CRYSTA Stress Viewer software package (Photron Co., Ltd.). This background-subtraction process is based on the presence of a certain degree of unevenness on the plate surface or birefringence in the plate or stage itself, which can cause a spatial distribution of retardation even when there is no flow. The position of ROI-A was chosen to avoid possible interference at the edges of the plate by the liquid--air interface or the liquid meniscus\cite{Ewoldt2015a}. 
ROI-A was also chosen to keep the uniformity of the lighting system, as the polarized image sensors use the light intensity itself to calculate the birefringence, as explained in the previous section. Another arc-shaped region, ROI-B (shown in red in inset~2 of Fig.~\ref{fig:Setup}b), with a width of 10~pixels (440~$\upmu$m), was also chosen for quantitative characterization of the flow birefringence. This was selected to ensure that the shear-rate variation along the radial direction was kept to a minimum, with $\approx$0.8\% difference. Therefore, within ROI-B, the shear rate applied to the microscopic fluid is $\dot\gamma_{\rm rep} = 0.75R_0{\rm \Omega}/H$, and this was considered as the representative shear rate in the polarization measurements. Here, $\rm \Omega$~[rad/s] is the angular velocity of the rotating plate.

The parallelism of the plate to the stage was not perfect. Therefore, due to changes in the integral of the birefringence, the measured retardation also changed with time. To exclude this effect, measurements were taken after a sufficient time had passed from the start of the plate's rotation and averaged over 1000~frames. As the polarization cameras measure the light intensity distribution as an image, minute noise in the intensity values is included in the calculation results as a constant retardation value, even if the fluid is not flowing. Figure~\ref{fig:Setup}c shows the time variation of the spatial average retardation in ROI-B measured with no flow. The retardation at a certain level was measured to be 1.46~nm; therefore, the birefringence was calculated on the assumption that this value ($\delta_{n, \rm error}\approx1.5\times10^{-5}$) was included as a measurement error or uncertainty.

\subsection*{Working fluid: Cellulose nanocrystal suspensions}
\begin{figure}[h]
\begin{center}
\includegraphics[width=0.6\textwidth]{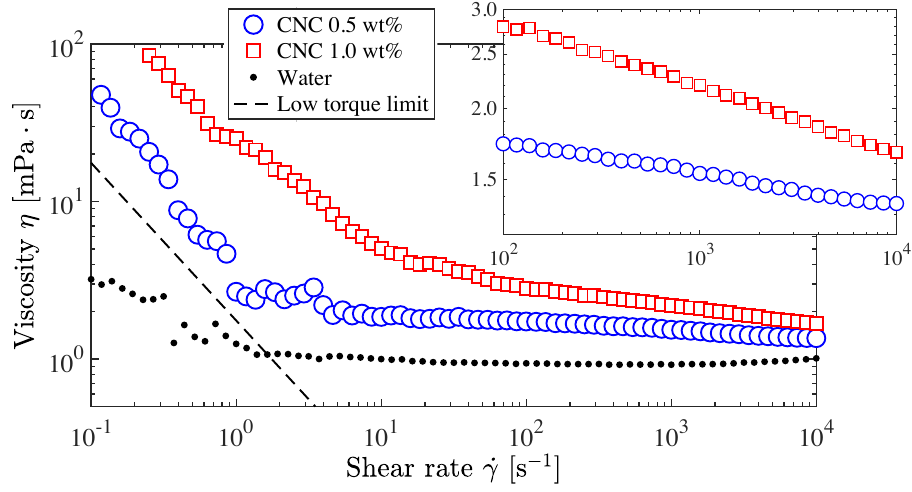}
\caption{Steady shear viscosity $\eta$ for CNC suspensions at different concentrations. 
The inset shows an enlarged region at~$\dot\gamma = 10^2$–$10^4$~s$^{-1}$.}
\label{fig:shear_curve}
\end{center}
\end{figure}

Suspensions of CNCs (Alberta Pacific Co., Ltd.) of two different concentrations were studied: 0.5~wt\% and 1.0~wt\%. 
These suspensions were prepared by mixing CNCs with ultrapure water using a hot stirrer at 40$^\circ$C and 650~rpm for more than 24~hours. 
The shear viscosities of these prepared CNC suspensions were then measured using a rheometer equipped with a 50-mm-diameter cone plate (CP50–0.5, Anton Paar Co., Ltd.) to obtain $\eta$ values. The measurement results are shown in Fig.~\ref{fig:shear_curve}. 
As can be seen, the 0.5~wt\% CNC suspension behaved like a Newtonian fluid, whereas the 1.0~wt\% CNC suspension showed a weak shear-thinning nature. 
Here, the clouding number $N$ is introduced to note the particle-particle interaction of the CNCs.
$N$ is expressed as $N=2c_v(l/d)^2/3$~[–] using the particle volume concentration $c_v$ (can be estimated from the CNC's density), particle length $l$ and particle width $d$\cite{Kerekes1992}.
For $N<1$, particles are relatively free to move; conversely, collisions between particles occur for $N>1$.
The aspect ratio of the CNCs used in the present study is comparable to those used in previous studies ($l/d\approx 16$)\cite{Lane2023, Lane2022, Shafiei-Sabet2012}.
Referring to morphology and the CNC density of 1,500~$\rm kg/m^3$\cite{Calabrese2021}, the $N$ for 0.5 wt\% and 1.0 wt\% suspension was 0.56 and 1.13, respectively.
$N>1$ may be one of the reasons for the non-Newtonian nature of the 1.0wt\% CNC suspension as shown in Fig.~\ref{fig:shear_curve}.
However, the change the shear viscosity $\eta$ was only $O(0.1)$~mPa$\cdot$s. for changes in shear rate $\dot\gamma$ between $10^3$–$10^4$~[s$^{-1}$].
In addition, the normal stress (Weizenberg effect) was also measured using a rheometer, and the values were small enough to be regarded as measurement errors.
Consequently, in the present study, we regarded the dilute CNC suspensions as Newtonian fluids. 

\section*{Results and discussion}
\begin{figure}[tb!]
\begin{center}
\includegraphics[width=1.0\textwidth]{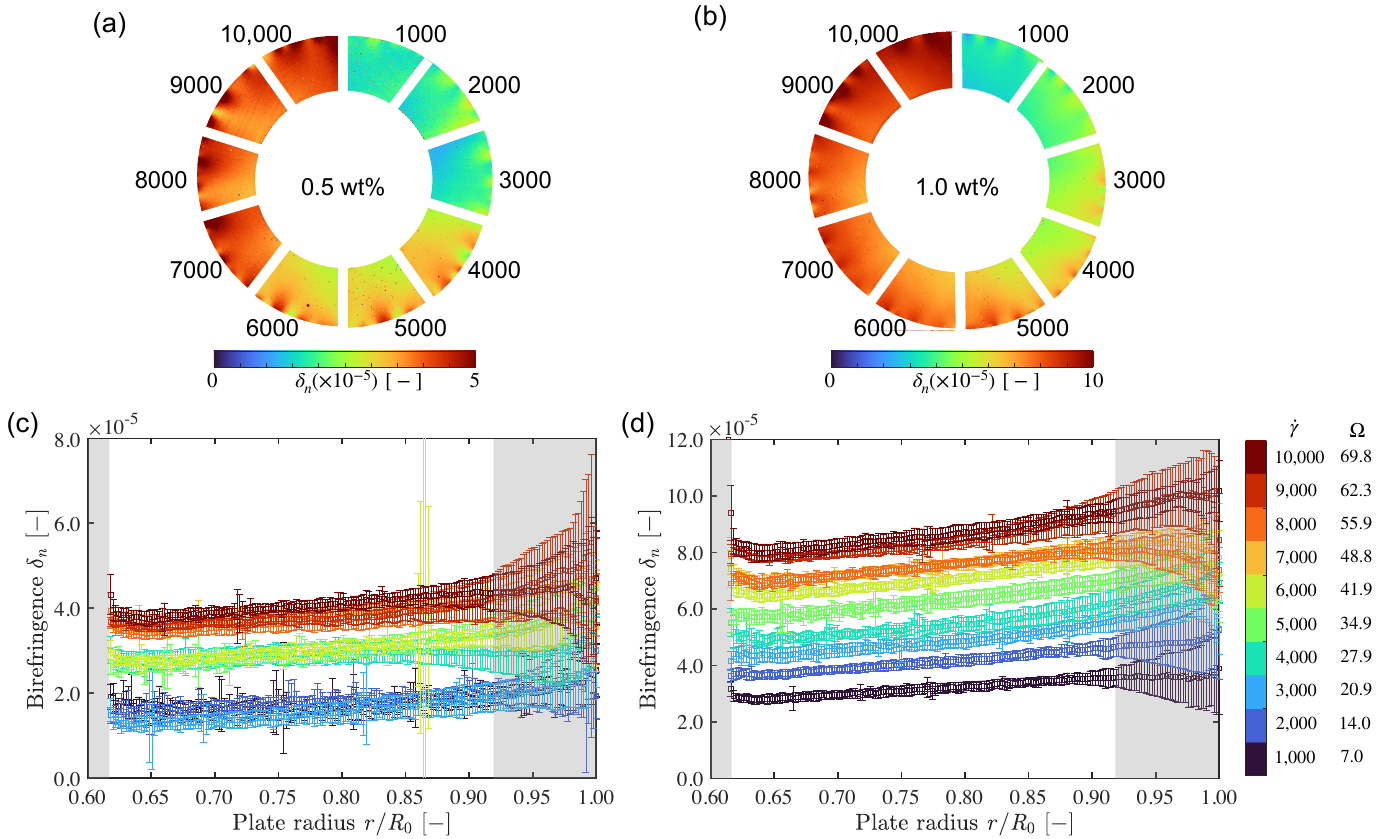}
\caption{Visualized birefringence fields within ROI-A for (a)~0.5~wt\% and (b)~1.0~wt\% CNC suspensions under steady-state conditions. Radial birefringence distributions of the plate in (c)~0.5 wt\% and (d)~1.0 wt\% CNC suspensions, in which the grey shaded areas correspond to ROI-A. The error bars show the standard deviations and all data were time and circumferential averaged.}
\label{fig:Result:Birefringence}
\end{center}
\end{figure}

Visualized birefringence fields at different shear rates are shown in Fig.~\ref{fig:Result:Birefringence}a and Fig.~\ref{fig:Result:Birefringence}b. 
As can be seen, the birefringence increased significantly as the shear rate was increased. This is because the CNCs attain more uniform orientations with increasingly strong shear leading to a stronger optical-anisotropy.
The magnitude of birefringence obtained was $\delta_n \sim O(10^{-5})$, comparable to the values found in previous studies\cite{Decruppe1995,Ito2016,Lane2021,Lane2022}.
In the experiments, we observed an uneven distribution of birefringence outside ROI-A, which was seen to vary unsteadily. 
This was considered to be an effect of interference at the liquid--air interface due to the plate's rotation.

We now consider an analytical model to help understand the measurement results. 
A simple analytical model of shear flow between the plates can be expressed:
\begin{equation}
{\bf u}(x,y,z) = {\rm \Omega}\frac{z}{H}[-y,x,0]^\mathrm{T},
\label{eq:analytical}
\end{equation}
where $H$ is the gap height. The strain tensor $\dot e$ is:
\begin{equation}
    \dot{e}  =\frac{{\rm \Omega}}{H}
    \begin{bmatrix}
    0 & 0 & -y \\
    0 & 0 & x \\
    -y& x & 0 \\
    \end{bmatrix}.
\label{eq:tensor}
\end{equation}
This means that the birefringence measured in the present study was induced by strain (stress) components along the optical axis. 
Substituting each component of the above tensor into Eqs.~(\ref{eq:sol-A}) and (\ref{eq:sol-B}) yields:
\begin{equation}
\delta_n= (C_2 \eta^2){\rm \Omega}^2\frac{x^2+y^2}{H^2} = (C_2 \eta^2)~\left(\frac{r\rm \Omega}{H}\right)^2.
\label{eq:subtensor}
\end{equation}
Thus, the birefringence is particularly dependent on the square of the radius $r$ and the square of the plate's angular velocity $\rm \Omega$, as long as $C_2$ and $\eta$ are constant.  

Line profiles of birefringence distributed across the radial direction of ROI-A at varying shear rates are shown in Fig.~\ref{fig:Result:Birefringence}c and Fig.~\ref{fig:Result:Birefringence}d. 
These show that the shear rate increases outwardly from the centre of the plate, which leads to an increase in the birefringence. 
When these results are compared with the analytical model, the quadratic dependence of the birefringence on $r$ is evident in the experimental results; however, the birefringence is clearly not proportional to the square of $\rm \Omega$. 
This deviation indicates that $C_2$ is not a constant value, and in particular that it varies with the shear rate. 
Details are given later in this section.

To further investigate the variation of the stress-optic coefficient $C_2$, its magnitude was estimated and compared. 
From Eq.~(\ref{eq:subtensor}), the magnitudes of each parameter are $\delta_n \sim O(10^{-5})$, $\eta \sim O(1)$~mPa$\cdot$s, $r\sim O(10)$~mm, ${\rm \Omega} \sim O(10^1$–$10^2)$~rad/s, and $H\sim O(100)~\upmu$m. By magnitude comparison, the magnitude of $C_2$ was estimated to be $C_2\sim O(10^{-7}$--$10^{-6})$~Pa$^{-2}$. 
To the best of our knowledge, this is the first systematic measurement report to identify the $C_2$ value of a birefringent fluid.
The stress-optic coefficient $C_1$, generally known as the photoelastic modulus, has been widely investigated in solid polymers, and it is known to have a magnitude of $O(10^{-12})$~Pa$^{-1}$\cite{Koike2004}. 
In contrast, there have been a few reports on investigations of the $C_1$ values of fluids. 
For reference, the $C_1$ values obtained for fluids are: worm-like micelles $O(10^{-7})$~Pa$^{-1}$\cite{Shikata1994,Ober2011,Ito2016}, aqueous xanthan gum solutions (0.3--0.7~mg/cm$^3$) $3.3\times 10^{-8}$~Pa$^{-1}$\cite{Yevlampieva1999}, and a 0.5~wt\% CNC suspension $O(10^{-5})$~Pa$^{-1}$\cite{Nakamine2024}.
Although simple comparisons are difficult because of the different units, i.e., Pa$^{-1}$ and Pa$^{-2}$, the magnitude of $C_2$ was found to be smaller than that of $C_1$.
Nevertheless, it is inappropriate to assume that $C_2$ can be ignored in the presence of stress distribution along the optical axis as discussed in the Introduction of this paper.

\begin{figure}[tb!]
\begin{center}
\includegraphics[width=0.65\textwidth]{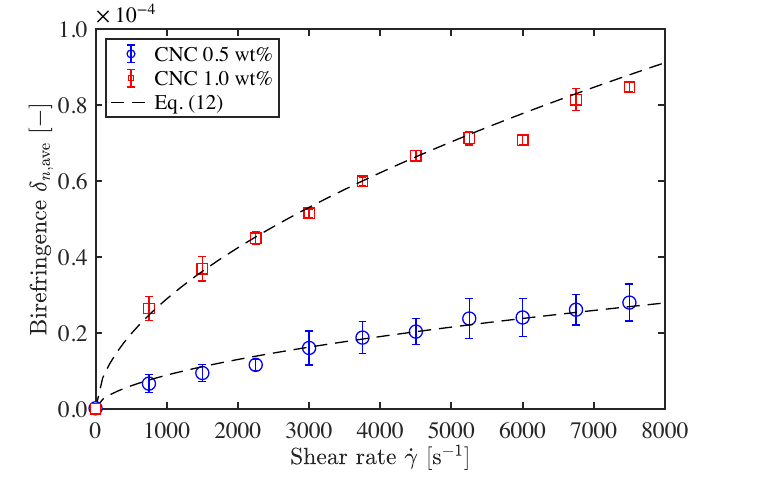}
\caption{Spatiotemporally averaged birefringence $\delta_{n,\rm ave}$ and corresponding fit using Eq.~(\ref{eq:powarlaw}) with $A=9.40\times 10^{-6}$~$\rm $, $n=0.552$ and $m = 1.71$, in which the error bars represent the standard deviation.}
\label{fig:Gammmavsdelta}
\end{center}
\end{figure}

Next, the trend of birefringence with respect to the shear rate was investigated. 
In Fig.~\ref{fig:Gammmavsdelta}, the vertical axis shows the spatiotemporally averaged birefringence $\delta_{n,\rm ave}$ in ROI-B, while the horizontal axis shows the representative shear rate $\dot\gamma_{\rm rep}$. 
It can be seen that the birefringence is increasing exponentially.
When modelling the relationship between the flow birefringence and shear rate, the following empirical nonlinear model was proposed by Lane et~al.\cite{Lane2022}:
\begin{equation}
    \delta_n=(A\cdot \dot\gamma)^{n} \cdot c^{m},
    \label{eq:powarlaw}
\end{equation}
where $c$~[--] is the concentration of the suspension, and $A$~[s], $n$~[--], and $m$~[--] are fitting parameters. 
The experimental results were fitted using Eq.~(\ref{eq:powarlaw}), and the black dash-dotted lines in Fig.~\ref{fig:Gammmavsdelta} show the results. 
The fitting parameters were $A=9.40\times 10^{-6}$~s, $n=0.552$ and $m = 1.71$.
It should be emphasized that this model is based on the results of polarization measurements conducted from the vorticity direction to the shear using a CC-type rheometer, which is different from that used in the present study.

\begin{figure}[tb!]
\begin{center}
\includegraphics[width=0.6\textwidth]{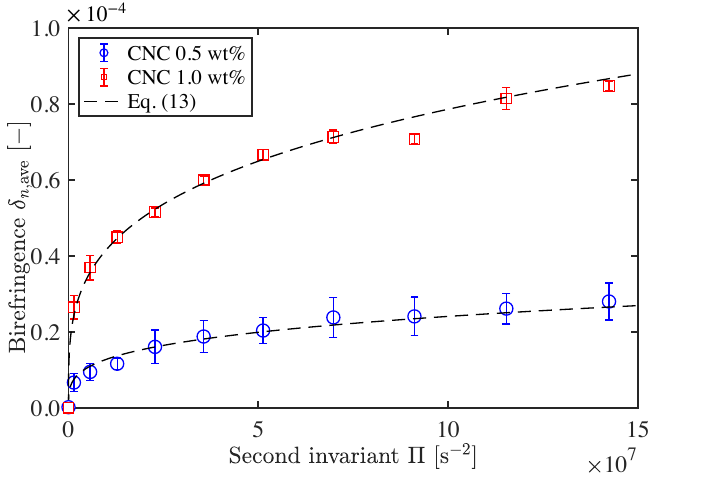}
\caption{Spatiotemporally averaged birefringence $\delta_{n,\rm ave}$ and corresponding fit using Eq.~(\ref{eq:powarlaw_new}) with $A=2.94\times 10^{-9}$~$\rm $, $\alpha=0.276$ and $\beta = 1.71$, in which the error bars represent the standard deviation.}
\label{fig:Main}
\end{center}
\end{figure}

As described in the previous section, flow birefringence is induced by the aligned orientation of crystals or polymer chains dispersed in a fluid. 
For generalization, and to provide a better prospect for the physical interpretation of the present experimental results, we attempt to disentangle the birefringence $\delta_n$ from the suspension concentration $c$ and the direction of the polarization measurement from the effect of CNC alignment due to the flow. 
We used tensor invariants that are independent of the coordinate system.
Therefore, the discussion henceforth will be based on the second invariant of the deformation-rate tensor, $\rm \Pi$~[s$^{-2}$], i.e.,
\begin{equation}
    \delta_{n}=(B\cdot {\rm \Pi})^{\alpha} \cdot c^{\beta}.
    \label{eq:powarlaw_new}
\end{equation}
Here, $B$~[s$^{2}$], $\alpha$~[--], and $\beta$~[--] are defined as new parameters. 
In the present system, $\Pi$ can be calculated as:
\begin{equation}
{\rm \Pi}({\bf E})= \sum\limits_{i=1}^3\sum\limits_{j=1}^3 {\bf E}_{ij}{\bf E}_{ji}=\frac{1}{2}\left(\frac{\rm \Omega}{H}\right)^2(x^2+y^2)=\frac{9}{8}{\dot\gamma_{\rm rep}}^2,
\label{eq:inavariant}
\end{equation}
where ${\bf E}$ is the deformation-rate tensor:
\begin{equation}
   {\bf E} = \frac{1}{2}\lbrace(\nabla {\bf u})+(\nabla {\bf u})^\mathrm{T}\rbrace.
\end{equation}
Since $\rm \Pi$ is proportional to the square of the shear rate $\dot\gamma$ from Eq.~(\ref{eq:inavariant}), $n = 2\alpha$ holds here.
The $\delta_n$ measurements were fitted using Eq.~(\ref{eq:powarlaw_new}), and the results are shown by the black dashed lines in Fig.~\ref{fig:Main}. 
The fitting parameters were $B = 2.94\times 10^{-9}$~s$^2$, $\alpha = 0.276$~($n = 0.552$), and $\beta = 1.71$.
Also represented in Fig.~\ref{fig:Gammmavsdelta}, it has been well reported that birefringence generally increases nonlinearly as the shear rate is increased\cite{Santos2023,Lane2022,Decruppe1995,Peebles1964}. 
This trend remains unchanged when organized by invariant, and the increase in birefringence per invariant decreases with decreasing gradient at higher invariants.

We focus our discussion on the exponent which characterizes the tendency of birefringence with the deformation of the fluid~(solvent).
Calabrese et~al.\cite{Calabrese2021} studied the birefringence of 0.1~wt\% CNC suspensions, and they proposed a proportionality of $\delta_n\propto{\dot\gamma}^{0.9}$ from the results of their experiments. They reported CNC geometries with an average length of 260~nm and an average width of 4.8~nm. 
Note that the 0.1~wt\% CNC suspension was stated to be a dilute region with no particle interactions. Lane et~al. also found the relationship $\delta_n\propto{\dot\gamma}^{0.537}$ in part of an investigation into whether CNC suspensions (0.7--1.3~wt\%) could be used in studies of flow birefringence\cite{Lane2022}. 
These suspensions were at concentrations above the dilute region, where particle interactions need to be considered. 
The exponent of 0.9 differs significantly from the present results, while 0.537 is remarkably close. We assume that the difference in $n$ results from different CNC particle-interaction behaviours.
In the existing literature, the relationship between CNC particle shear alignment and rotational diffusion is described by the P\'{e}clet number $Pe=|{\rm E}|/D_r$, where $|{\rm E}|$ is the characteristic deformation rate\cite{Doi1988}. 
When $Pe\geq1$, convective forces are strong enough to align CNCs to the flow direction, eventually inducing birefringence. 
Here, $D_r$ is the rotational diffusion coefficient which plays an important role in flow birefringence.
The $D_r$ is determined by the CNC rod length $l$ and the suspension concentration $c$, giving $D_r\propto c^{-2}l^{-9}$\cite{Doi1988,Maguire1980}.
This indicates that the optical properties of CNC suspensions are due to different rod lengths and concentrations, resulting in different exponents.
The present experimental results and those of Lane et~al.\cite{Lane2022} indicate that regardless of the direction of polarization measurement with respect to shear, there seems to be a common physical background that leads to the flow birefringence following a power law.

\begin{figure}[tb!]
\begin{center}
\includegraphics[width=0.65\textwidth]{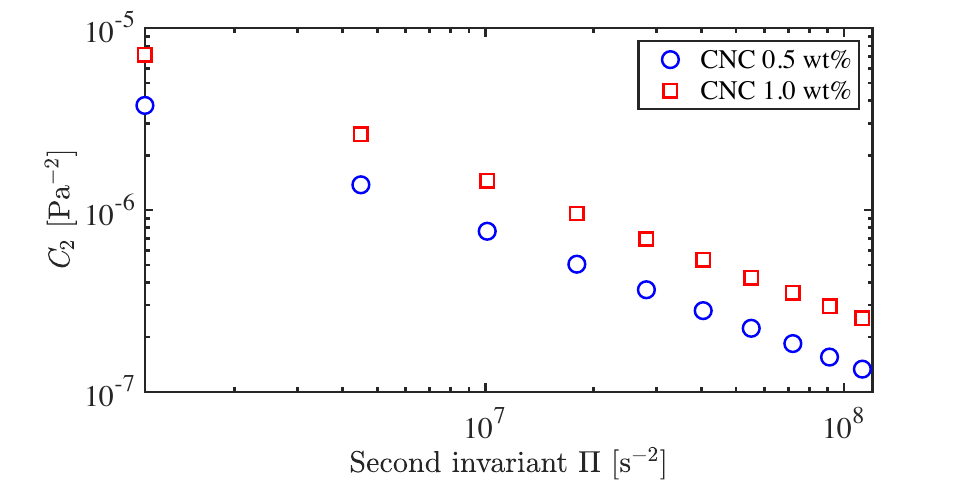}
\caption{Determined values of $C_2$ for each CNC suspension in base-10 logarithmic scale calculated from Eq.~(\ref{eq:C2Pi}).}
\label{fig:C2}
\end{center}
\end{figure}

The consistency between the SOL (Eq.~(\ref{eq:subtensor})) and the empirical relation (Eq.~(\ref{eq:powarlaw_new})) is now discussed. Since both equations are equally related by $\delta_n$,
\begin{equation}
C_2\eta^2\dot\gamma^2 = (B\cdot {\rm \Pi})^\alpha \cdot c^\beta \rightarrow C_2 = \frac{B^\alpha c^{\beta}}{\eta^2}{\rm \Pi}^{\alpha-1}.
\label{eq:C2Pi1}
\end{equation}
If we assume that the CNC suspension in the present study behaves as a Newtonian fluid, i.e., the change in the viscosity coefficient $\eta$ with the shear rate is sufficiently small,
\begin{equation}
    C_2\propto B^\alpha c^\beta{\rm \Pi}^{\alpha-1}=(4.4\times 10^{-3}) c^{\beta} {\rm \Pi}^{-0.724}.
\label{eq:C2Pi}
\end{equation}
From Eq.~(\ref{eq:C2Pi}), the coordinate-independent invariant ${\rm \Pi}$ and the pre-factor $B$, which reflects the direction of polarization measurement, along with the concentration $c$, disentangle the magnitude of the stress-optic coefficient $C_2$, which depends on the coordinate. In other words, birefringence (optical-anisotropy) can be universally described by an invariant and certain kinds of biases. The specific values of $C_2$ in the present study are given in Fig.~\ref{fig:C2} using the infinite-shear viscosity at $\dot\gamma = 10^4$~s$^{-1}$ with 1.3 and 1.7~mPa$\cdot$s for 0.5~wt\% and 1.0~wt\% CNC suspensions, respectively. 
These values are consistent with the range of the earlier magnitude-approximation result. Additionally, $C_2$ was found to increase with increasing concentration of CNC suspension. This means that the sensitivity to stress (degree of optical-anisotropy) increased with increasing concentration, which is a well-known trend.

In summary, as in the PP-type rheometer used in the present experiments, $C_2$ can be calibrated by simply using a flow field with a shear-velocity distribution along the optical axis. The polarization measurement direction-dependent pre-factor $B$ is then incorporated into the SOL as $C_2$. 
We suggest that $C_2$ as obtained in the present method can be replaced by $C_2=f(B,~\rm \Pi,~\it c)$ in the SOL to give a good expression for the optical-anisotropy due to the stress distribution along the optical axis.
A detailed study of the physical quantities dominating the pre-factor $B$, which determines $C_2$, is a subject for future work.
We believe that a systematic investigation of $B$ has the potential to contribute to our understanding of the flow dynamics of complex fluids.

\section*{Conclusion}
This paper provides the results of an investigation into the effect of the stress distribution along the camera's optical axis on optical-anisotropy. 
In the experiments, rheo-optical measurements were performed using a PP-type rheometer and a polarization camera for a fundamental understanding of SOL. 
For the birefringent fluid, a dilute aqueous CNC suspension was used, showing properties close to those of a Newtonian fluid.
The birefringence was found to be induced by the shear-stress distribution along the camera's optical axis, rather than by a principal stress difference as considered in conventional photoelastic theory in a 2D stress field. 
The representative birefringence was found to increase nonlinearly and monotonically with the shear rate, following a power function of the second invariant of the deformation-rate tensor. In addition, from the well-known SOL and an empirical equation proposed in a previous study, the contribution of stress to optical-anisotropy can be determined by a pre-factor representing the direction of polarization measurement and the invariant. The degree of optical-anisotropy (birefringence) in the present study corresponds to the coefficient (stress-optic coefficient $C_2$) of the nonlinear term of the SOL. Based on the present experimental results, it can be said that the assumption that $C_2=0$ is not appropriate, especially when the SOL is applied to 3D fluid stress fields in which the stress is distributed along the camera's optical axis. By defining $C_2$ as a non-zero invariant function, the contribution of the stress component along the camera's optical axis in each experimental system can be quantitatively described.
In the future, SOL with $C_2$ term may provide a deeper understanding of complex fluids showing non-Newtonian properties such as particle-particle interactions and polymer extension and contraction.

\appendix
\def\thesection{Appendix}
\section{Estimation of experimental window}\label{sec:window}
In rheological measurements of biomaterials, the results may not always show the real behaviour of the liquid due to various disturbances. Possible disturbance factors include instrument specifications, secondary flows at high velocity\cite{Sdougos1984}, and surface-tension forces\cite{Ewoldt2015a,Johnston2013}. To correctly understand the rheological and associated polarization measurements, it is important to understand these limitations. To identify limitations, an experimental window was considered\cite{Ewoldt2015a}. We assumed that within the delineated region, there were no artefacts in the rheological and associated polarization measurements.

\begin{figure}[tb!]
\begin{center}
\includegraphics[width=0.85\textwidth]{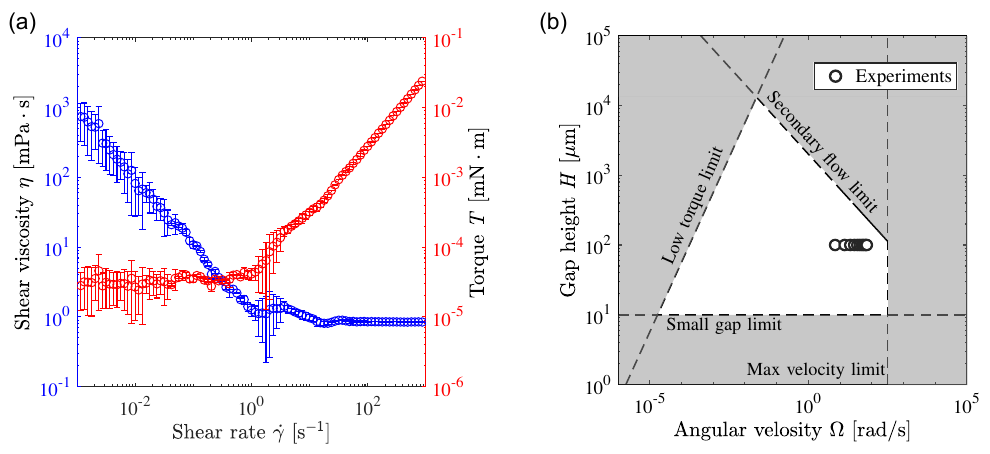}
\caption{(a)~Results of shear-viscosity measurements of water and the torque transition during the measurements. The torque can be seen to plateau up to 35~nN$\cdot$m, which we defined as $T_{\rm min}$. (b)~Experimental window for the rheometer used in the present study. The representative values used were $H_{\rm min}= 10$~$\upmu$m, $\rm \Omega_{max} = 314$~rad/s, $Re_{\rm max} = 4.0$, $\eta = 1.0$~mPa$\cdot$s, and $\rho = 1000$~kg/m$^3$. The open circles show the present experimental conditions.}
\label{fig:Limitation}
\end{center}
\end{figure}

First, the minimum torque $T_{\rm min}$ was obtained experimentally rather than from the equipment specifications. This is because its value might be larger than the equipment specifications due to the surface tension of the fluid\cite{Soulages2009,RODD2005}. To estimate the minimum torque $T_{\rm min}$ of the present rheometer, Fig.~\ref{fig:Limitation}a shows the results of viscosity measurements from a low shear rate with water. The torque showed an almost constant value at low shear rates, thereby providing misleading rheological results, in which the water shows shear-thinning viscosity. In contrast, at shear rates above 1~s$^{-1}$, the torque measurements varied with the applied shear rate. We therefore defined the minimum torque as the value at the beginning of the increase. From Fig.~\ref{fig:Limitation}a, we can determine the minimum torque $T_{\rm min}$ to be 35~nN$\cdot$m. Based on the criterion $T>T_{\rm min}$, where the torque from steady viscosity is $T=(\eta {\rm \Omega} / H)/F_\tau$, the boundary line is defined by:
\begin{equation}
    H<\frac{\eta {\rm\Omega}}{F_\tau T_{\rm min}}.
    \label{eq:low tprque}
\end{equation}
Here, $\rm \Omega$~[rad/s] is the angular velocity and $\eta$ is the viscosity of the fluid. 
This equation is than used in Fig.~\ref{fig:Limitation}b for the parallel-plate geometry $F_{\tau}=3/(2\pi {R_0}^3)$, where $R_0$ is the plate radius\cite{Ewoldt2015a}. The equipment limit for maximum speed is simply determined by the criterion:
\begin{equation}
    {\rm \Omega}<{\rm \Omega}_{\rm max},
     \label{eq:velosity}
\end{equation}
where $\rm \Omega_{\rm max}$~[rad/s] is the maximum rotation speed of the motor equipped in the present rheometer. The limit of secondary flow, the maximum Reynolds number $ Re_{\rm max}$, sets the boundary line, and this is based on the definition $Re = \rho {\rm \Omega} H^2/ \eta$ for parallel plates:
\begin{equation}
    H<\left(\frac{Re_{\rm max}\eta}{\rho \rm\Omega}\right)^{1/2}.
     \label{eq:secondary flow}
\end{equation}
Here, $\rho$~[kg/m$^3$] is the density of the fluid. In the present study, we set $Re_{\rm max}=4$, which is a criterion within a torque measurement error of 1\% according to Ewoldt et~al.\cite{Ewoldt2015a}. Finally, a small-gap limit was set by applying the minimum height that can be set, which is $H_{\rm min} = 10~\upmu$m:
\begin{equation}
    H > H_{\rm min}.
    \label{eq:smallgap}
\end{equation}
By plotting the boundaries defined by Eqs.~(\ref{eq:low tprque})--(\ref{eq:smallgap}), we obtain the experimental window shown in Fig.~\ref{fig:Limitation}b.

\bibliography{Ref}

\begin{thebibliography}{10}
\urlstyle{rm}
\expandafter\ifx\csname url\endcsname\relax
  \def\url#1{\texttt{#1}}\fi
\expandafter\ifx\csname urlprefix\endcsname\relax\def\urlprefix{URL }\fi
\expandafter\ifx\csname doiprefix\endcsname\relax\def\doiprefix{DOI: }\fi
\providecommand{\bibinfo}[2]{#2}
\providecommand{\eprint}[2][]{\url{#2}}

\bibitem{Adamczyk1988}
\bibinfo{author}{Adamczyk, A.~A.} \& \bibinfo{author}{Rimai, L.}
\newblock \bibinfo{journal}{\bibinfo{title}{2-dimensional particle tracking velocimetry ({{PTV}}): {{Technique}} and image processing algorithms}}.
\newblock {\emph{\JournalTitle{Experiments in Fluids}}} \textbf{\bibinfo{volume}{6}}, \bibinfo{pages}{373--380}, \doiprefix\url{https://doi.org/10.1007/BF00196482} (\bibinfo{year}{1988}).

\bibitem{Oudheusden2013}
\bibinfo{author}{van Oudheusden, B.~W.}
\newblock \bibinfo{journal}{\bibinfo{title}{{{PIV-based}} pressure measurement}}.
\newblock {\emph{\JournalTitle{Measurement Science and Technology}}} \textbf{\bibinfo{volume}{24}}, \bibinfo{pages}{032001}, \doiprefix\url{https://doi.org/10.1088/0957-0233/24/3/032001} (\bibinfo{year}{2013}).

\bibitem{McAfee1974}
\bibinfo{author}{McAfee, W.~J.} \& \bibinfo{author}{Pih, H.}
\newblock \bibinfo{journal}{\bibinfo{title}{Scattered-light flow-optic relations adaptable to three-dimensional flow birefringence}}.
\newblock {\emph{\JournalTitle{Experimental Mechanics}}} \textbf{\bibinfo{volume}{14}}, \bibinfo{pages}{385--391}, \doiprefix\url{10.1007/BF02324941} (\bibinfo{year}{1974}).

\bibitem{Martins1986}
\bibinfo{author}{Martins, A.~F.}, \bibinfo{author}{Esnault, P.} \& \bibinfo{author}{Volino, F.}
\newblock \bibinfo{journal}{\bibinfo{title}{Measurement of the viscoelastic coefficients of main-chain nematic polymers by an {NMR} technique}}.
\newblock {\emph{\JournalTitle{Physical Review Letters}}} \textbf{\bibinfo{volume}{57}}, \bibinfo{pages}{1745--1748}, \doiprefix\url{https://doi.org/10.1103/PhysRevLett.57.1745} (\bibinfo{year}{1986}).

\bibitem{Odagiri2016}
\bibinfo{author}{Odagiri, K.} \emph{et~al.}
\newblock \bibinfo{journal}{\bibinfo{title}{Non-invasive evaluation of pulmonary arterial blood flow and wall shear stress in pulmonary arterial hypertension with {{3D}} phase contrast magnetic resonance imaging}}.
\newblock {\emph{\JournalTitle{SpringerPlus}}} \textbf{\bibinfo{volume}{5}}, \bibinfo{pages}{1071}, \doiprefix\url{10.1186/s40064-016-2755-7} (\bibinfo{year}{2016}).

\bibitem{Aben2012}
\bibinfo{author}{Aben, H.} \& \bibinfo{author}{Guillemet, C.}
\newblock \emph{\bibinfo{title}{Photoelasticity of Glass}} (\bibinfo{publisher}{{Springer Science \& Business Media}}, \bibinfo{year}{2012}).

\bibitem{Lautre2015}
\bibinfo{author}{Lautre, N.~K.}, \bibinfo{author}{Sharma, A.~K.}, \bibinfo{author}{Kumar, P.} \& \bibinfo{author}{Das, S.}
\newblock \bibinfo{journal}{\bibinfo{title}{A photoelasticity approach for characterization of defects in microwave drilling of soda lime glass}}.
\newblock {\emph{\JournalTitle{Journal of Materials Processing Technology}}} \textbf{\bibinfo{volume}{225}}, \bibinfo{pages}{151--161}, \doiprefix\url{https://doi.org/10.1016/j.jmatprotec.2015.05.026} (\bibinfo{year}{2015}).

\bibitem{Ramesh2021}
\bibinfo{author}{Ramesh, K.}
\newblock \emph{\bibinfo{title}{Developments in Photoelasticity: {A} Renaissance}} (\bibinfo{publisher}{{IOP Publishing}}, \bibinfo{year}{2021}).

\bibitem{Prabhakaran1975}
\bibinfo{author}{Prabhakaran, R.}
\newblock \bibinfo{journal}{\bibinfo{title}{On the stress-optic law for orthotropic-model materials in biaxial-stress fields}}.
\newblock {\emph{\JournalTitle{Experimental Mechanics}}} \textbf{\bibinfo{volume}{15}}, \bibinfo{pages}{29--34}, \doiprefix\url{https://doi.org/10.1007/BF02318522} (\bibinfo{year}{1975}).

\bibitem{Doyle1978}
\bibinfo{author}{Doyle, J.~F.} \& \bibinfo{author}{Danyluk, H.~T.}
\newblock \bibinfo{journal}{\bibinfo{title}{Integrated photoelasticity for axisymmetric problems}}.
\newblock {\emph{\JournalTitle{Experimental Mechanics}}} \textbf{\bibinfo{volume}{18}}, \bibinfo{pages}{215--220}, \doiprefix\url{https://doi.org/10.1007/BF02328416} (\bibinfo{year}{1978}).

\bibitem{Sampson1970}
\bibinfo{author}{Sampson, R.~C.}
\newblock \bibinfo{journal}{\bibinfo{title}{A stress-optic law for photoelastic analysis of orthotropic composites}}.
\newblock {\emph{\JournalTitle{Experimental Mechanics}}} \textbf{\bibinfo{volume}{10}}, \bibinfo{pages}{210--215}, \doiprefix\url{https://doi.org/10.1007/BF02324034} (\bibinfo{year}{1970}).

\bibitem{Srinath1974}
\bibinfo{author}{Srinath, L.~S.} \& \bibinfo{author}{Sarma, A. V. S. S. S.~R.}
\newblock \bibinfo{journal}{\bibinfo{title}{Determination of the optically equivalent model in three-dimensional photoelasticity}}.
\newblock {\emph{\JournalTitle{Experimental Mechanics}}} \textbf{\bibinfo{volume}{14}}, \bibinfo{pages}{118--122}, \doiprefix\url{https://doi.org/10.1007/BF02324775} (\bibinfo{year}{1974}).

\bibitem{Yokoyama2023}
\bibinfo{author}{Yokoyama, Y.} \emph{et~al.}
\newblock \bibinfo{journal}{\bibinfo{title}{Integrated photoelasticity in a soft material: Phase retardation, azimuthal angle, and stress-optic coefficient}}.
\newblock {\emph{\JournalTitle{Optics and Lasers in Engineering}}} \textbf{\bibinfo{volume}{161}}, \bibinfo{pages}{107335}, \doiprefix\url{https://doi.org/10.1016/j.optlaseng.2022.107335} (\bibinfo{year}{2023}).

\bibitem{Noto2020}
\bibinfo{author}{Noto, D.}, \bibinfo{author}{Tasaka, Y.}, \bibinfo{author}{Hitomi, J.} \& \bibinfo{author}{Murai, Y.}
\newblock \bibinfo{journal}{\bibinfo{title}{Applicability evaluation of the stress-optic law in {{Newtonian}} fluids toward stress field measurements}}.
\newblock {\emph{\JournalTitle{Physical Review Research}}} \textbf{\bibinfo{volume}{2}}, \bibinfo{pages}{043111}, \doiprefix\url{https://doi.org/10.1103/PhysRevResearch.2.043111} (\bibinfo{year}{2020}).

\bibitem{Lane2023}
\bibinfo{author}{Lane, C.}, \bibinfo{author}{Baumann, F.}, \bibinfo{author}{Rode, D.} \& \bibinfo{author}{R{\"o}sgen, T.}
\newblock \bibinfo{journal}{\bibinfo{title}{Two-dimensional strain rate imaging study using a polarization camera and birefringent aqueous cellulose nanocrystal suspensions}}.
\newblock {\emph{\JournalTitle{Experiments in Fluids}}} \textbf{\bibinfo{volume}{65}}, \bibinfo{pages}{8}, \doiprefix\url{https://doi.org/10.1007/s00348-023-03730-8} (\bibinfo{year}{2023}).

\bibitem{Clemeur2004}
\bibinfo{author}{Clemeur, N.}, \bibinfo{author}{Rutgers, R. P.~G.} \& \bibinfo{author}{Debbaut, B.}
\newblock \bibinfo{journal}{\bibinfo{title}{Numerical evaluation of three dimensional effects in planar flow birefringence}}.
\newblock {\emph{\JournalTitle{Journal of Non-Newtonian Fluid Mechanics}}} \textbf{\bibinfo{volume}{123}}, \bibinfo{pages}{105--120}, \doiprefix\url{https://doi.org/10.1016/j.jnnfm.2004.07.002} (\bibinfo{year}{2004}).

\bibitem{Kim2017}
\bibinfo{author}{Kim, J.} \emph{et~al.}
\newblock \bibinfo{journal}{\bibinfo{title}{Monitoring the orientation of rare-earth-doped nanorods for flow shear tomography}}.
\newblock {\emph{\JournalTitle{Nature Nanotechnology}}} \textbf{\bibinfo{volume}{12}}, \bibinfo{pages}{914--919}, \doiprefix\url{https://doi.org/10.1038/nnano.2017.111} (\bibinfo{year}{2017}).

\bibitem{Alizadehgiashi2018}
\bibinfo{author}{Alizadehgiashi, M.} \emph{et~al.}
\newblock \bibinfo{journal}{\bibinfo{title}{Shear-induced alignment of anisotropic nanoparticles in a single-droplet oscillatory microfluidic platform}}.
\newblock {\emph{\JournalTitle{ACS Publications}}} \textbf{\bibinfo{volume}{34}}, \bibinfo{pages}{322--330}, \doiprefix\url{https://doi.org/10.1021/acs.langmuir.7b03648} (\bibinfo{year}{2018}).

\bibitem{Nakamine2024}
\bibinfo{author}{Nakamine, K.}, \bibinfo{author}{Yokoyama, Y.}, \bibinfo{author}{Worby, W. K.~A.}, \bibinfo{author}{Muto, M.} \& \bibinfo{author}{Tagawa, Y.}
\newblock \bibinfo{journal}{\bibinfo{title}{Flow birefringence of cellulose nanocrystal suspensions in three-dimensional flow fields: Revisiting the stress-optic law}}.
\newblock {\emph{\JournalTitle{{arXiv preprint}}}} \doiprefix\url{https://doi.org/10.48550/arXiv.2402.16351} (\bibinfo{year}{2024}).

\bibitem{Lodge1956}
\bibinfo{author}{Lodge, A.~S.}
\newblock \bibinfo{journal}{\bibinfo{title}{A network theory of flow birefringence and stress in concentrated polymer solutions}}.
\newblock {\emph{\JournalTitle{Transactions of the Faraday Society}}} \textbf{\bibinfo{volume}{52}}, \bibinfo{pages}{120--130}, \doiprefix\url{https://doi.org/10.1039/TF9565200120} (\bibinfo{year}{1956}).

\bibitem{Philippoff1961}
\bibinfo{author}{Philippoff, W.}
\newblock \bibinfo{journal}{\bibinfo{title}{Stress-optical analysis of fluids}}.
\newblock {\emph{\JournalTitle{Rheologica Acta}}} \textbf{\bibinfo{volume}{1}}, \bibinfo{pages}{371--375}, \doiprefix\url{https://doi.org/10.1007/BF01989069} (\bibinfo{year}{1961}).

\bibitem{Rothstein2002}
\bibinfo{author}{Rothstein, J.~P.} \& \bibinfo{author}{McKinley, G.~H.}
\newblock \bibinfo{journal}{\bibinfo{title}{A comparison of the stress and birefringence growth of dilute, semi-dilute and concentrated polymer solutions in uniaxial extensional flows}}.
\newblock {\emph{\JournalTitle{Journal of Non-Newtonian Fluid Mechanics}}} \textbf{\bibinfo{volume}{108}}, \bibinfo{pages}{275--290}, \doiprefix\url{https://doi.org/10.1016/S0377-0257(02)00134-9} (\bibinfo{year}{2002}).

\bibitem{Muto2022}
\bibinfo{author}{Muto, M.} \& \bibinfo{author}{Tagawa, Y.}
\newblock \bibinfo{journal}{\bibinfo{title}{Unsteady rheo-optical measurements of uniaxially extending liquid polymers}}.
\newblock {\emph{\JournalTitle{{arXiv preprint}}}} \doiprefix\url{https://doi.org/10.48550/arXiv.2204.13450} (\bibinfo{year}{2022}).

\bibitem{Muto2024}
\bibinfo{author}{Muto, M.}, \bibinfo{author}{Yoshino, T.} \& \bibinfo{author}{Tamano, S.}
\newblock \bibinfo{title}{Simultaneous measurement of extensional stress and flow birefringence field for uniaxially extending worm-like micellar solutions}, \doiprefix\url{https://doi.org/10.48550/arXiv.2404.17643} (\bibinfo{year}{2024}).

\bibitem{Tanaka2018}
\bibinfo{author}{Tanaka, R.}, \bibinfo{author}{Li, S.}, \bibinfo{author}{Kashiwagi, Y.} \& \bibinfo{author}{Inoue, T.}
\newblock \bibinfo{journal}{\bibinfo{title}{A self-build apparatus for oscillatory flow birefringence measurements in a co-cylindrical geometry}}.
\newblock {\emph{\JournalTitle{Nihon Reoroji Gakkaishi}}} \textbf{\bibinfo{volume}{46}}, \bibinfo{pages}{221--226}, \doiprefix\url{https://doi.org/10.1678/rheology.46.221} (\bibinfo{year}{2018}).

\bibitem{Lane2022}
\bibinfo{author}{Lane, C.}, \bibinfo{author}{Rode, D.} \& \bibinfo{author}{R{\"o}sgen, T.}
\newblock \bibinfo{journal}{\bibinfo{title}{Birefringent properties of aqueous cellulose nanocrystal suspensions}}.
\newblock {\emph{\JournalTitle{Cellulose}}} \textbf{\bibinfo{volume}{29}}, \bibinfo{pages}{6093--6107}, \doiprefix\url{https://doi.org/10.1007/s10570-022-04646-y} (\bibinfo{year}{2022}).

\bibitem{Kadar2020}
\bibinfo{author}{K{\'a}d{\'a}r, R.}, \bibinfo{author}{Fazilati, M.} \& \bibinfo{author}{Nypel{\"o}, T.}
\newblock \bibinfo{journal}{\bibinfo{title}{Unexpected microphase transitions in flow towards nematic order of cellulose nanocrystals}}.
\newblock {\emph{\JournalTitle{Cellulose}}} \textbf{\bibinfo{volume}{27}}, \bibinfo{pages}{2003--2014}, \doiprefix\url{https://doi.org/10.1007/s10570-019-02888-x} (\bibinfo{year}{2020}).

\bibitem{Detert2023}
\bibinfo{author}{Detert, M.}, \bibinfo{author}{Santos, T.~P.}, \bibinfo{author}{Shen, A.~Q.} \& \bibinfo{author}{Calabrese, V.}
\newblock \bibinfo{journal}{\bibinfo{title}{Alignment--rheology relationship of biosourced rod-like colloids and polymers under flow}}.
\newblock {\emph{\JournalTitle{Biomacromolecules}}} \textbf{\bibinfo{volume}{24}}, \bibinfo{pages}{3304--3312}, \doiprefix\url{https://doi.org/10.1021/acs.biomac.3c00347} (\bibinfo{year}{2023}).

\bibitem{Oba2016}
\bibinfo{author}{Oba, N.} \& \bibinfo{author}{Inoue, T.}
\newblock \bibinfo{journal}{\bibinfo{title}{An apparatus for birefringence and extinction angle distributions measurements in cone and plate geometry by polarization imaging method}}.
\newblock {\emph{\JournalTitle{Rheologica Acta}}} \textbf{\bibinfo{volume}{55}}, \bibinfo{pages}{699--708}, \doiprefix\url{https://doi.org/10.1007/s00397-016-0952-5} (\bibinfo{year}{2016}).

\bibitem{Mykhaylyk2016}
\bibinfo{author}{Mykhaylyk, O.~O.}, \bibinfo{author}{Warren, N.~J.}, \bibinfo{author}{Parnell, A.~J.}, \bibinfo{author}{Pfeifer, G.} \& \bibinfo{author}{Laeuger, J.}
\newblock \bibinfo{journal}{\bibinfo{title}{Applications of shear-induced polarized light imaging ({{SIPLI}}) technique for mechano-optical rheology of polymers and soft matter materials}}.
\newblock {\emph{\JournalTitle{Journal of Polymer Science Part B: Polymer Physics}}} \textbf{\bibinfo{volume}{54}}, \bibinfo{pages}{2151--2170}, \doiprefix\url{https://doi.org/10.1002/polb.24111} (\bibinfo{year}{2016}).

\bibitem{Hausmann2018}
\bibinfo{author}{Hausmann, M.~K.} \emph{et~al.}
\newblock \bibinfo{journal}{\bibinfo{title}{Dynamics of cellulose nanocrystal alignment during {3D} printing}}.
\newblock {\emph{\JournalTitle{ACS Nano}}} \textbf{\bibinfo{volume}{12}}, \bibinfo{pages}{6926--6937}, \doiprefix\url{https://doi.org/10.1021/acsnano.8b02366} (\bibinfo{year}{2018}).

\bibitem{Maxwell1874}
\bibinfo{author}{Maxwell, J.~C.}
\newblock \bibinfo{journal}{\bibinfo{title}{On double refraction in a viscous fluid in motion}}.
\newblock {\emph{\JournalTitle{Proceedings of the Royal Society of London}}} \textbf{\bibinfo{volume}{22}}, \bibinfo{pages}{46--47}, \doiprefix\url{https://doi.org/10.1098/rspl.1873.0011} (\bibinfo{year}{1874}).

\bibitem{Calabrese2021}
\bibinfo{author}{Calabrese, V.}, \bibinfo{author}{Haward, S.~J.} \& \bibinfo{author}{Shen, A.~Q.}
\newblock \bibinfo{journal}{\bibinfo{title}{Effects of shearing and extensional flows on the alignment of colloidal rods}}.
\newblock {\emph{\JournalTitle{Macromolecules}}} \textbf{\bibinfo{volume}{54}}, \bibinfo{pages}{4176--4185}, \doiprefix\url{https://doi.org/10.1021/acs.macromol.0c02155} (\bibinfo{year}{2021}).

\bibitem{Ober2011}
\bibinfo{author}{Ober, T.~J.}, \bibinfo{author}{Soulages, J.} \& \bibinfo{author}{McKinley, G.~H.}
\newblock \bibinfo{journal}{\bibinfo{title}{Spatially resolved quantitative rheo-optics of complex fluids in a microfluidic device}}.
\newblock {\emph{\JournalTitle{Journal of Rheology}}} \textbf{\bibinfo{volume}{55}}, \bibinfo{pages}{1127--1159}, \doiprefix\url{https://doi.org/10.1122/1.3606593} (\bibinfo{year}{2011}).

\bibitem{Yevlampieva1999}
\bibinfo{author}{Yevlampieva, N.~P.}, \bibinfo{author}{Pavlov, G.~M.} \& \bibinfo{author}{Rjumtsev, E.~I.}
\newblock \bibinfo{journal}{\bibinfo{title}{Flow birefringence of xanthan and other polysaccharide solutions}}.
\newblock {\emph{\JournalTitle{International Journal of Biological Macromolecules}}} \textbf{\bibinfo{volume}{26}}, \bibinfo{pages}{295--301}, \doiprefix\url{https://doi.org/10.1016/S0141-8130(99)00096-3} (\bibinfo{year}{1999}).

\bibitem{Doyle1982}
\bibinfo{author}{Doyle, J.~F.}
\newblock \bibinfo{journal}{\bibinfo{title}{On a nonlinearity in flow birefringence}}.
\newblock {\emph{\JournalTitle{Experimental Mechanics}}} \textbf{\bibinfo{volume}{22}}, \bibinfo{pages}{37--38}, \doiprefix\url{https://doi.org/10.1007/BF02325702} (\bibinfo{year}{1982}).

\bibitem{Aben1997}
\bibinfo{author}{Aben, H.} \& \bibinfo{author}{Puro, A.}
\newblock \bibinfo{journal}{\bibinfo{title}{Photoelastic tomography for three-dimensional flow birefringence studies}}.
\newblock {\emph{\JournalTitle{Inverse Problems}}} \textbf{\bibinfo{volume}{13}}, \bibinfo{pages}{215--221}, \doiprefix\url{https://doi.org/10.1088/0266-5611/13/2/002} (\bibinfo{year}{1997}).

\bibitem{Riera1969}
\bibinfo{author}{Riera, J.~D.} \& \bibinfo{author}{Mark, R.}
\newblock \bibinfo{journal}{\bibinfo{title}{The optical-rotation effect in photoelastic shell analysis}}.
\newblock {\emph{\JournalTitle{Experimental Mechanics}}} \textbf{\bibinfo{volume}{9}}, \bibinfo{pages}{9--16}, \doiprefix\url{https://doi.org/10.1007/BF02327872} (\bibinfo{year}{1969}).

\bibitem{Onuma2014}
\bibinfo{author}{Onuma, T.} \& \bibinfo{author}{Otani, Y.}
\newblock \bibinfo{journal}{\bibinfo{title}{A development of two-dimensional birefringence distribution measurement system with a sampling rate of 1.3{MHz}}}.
\newblock {\emph{\JournalTitle{Optics Communications}}} \textbf{\bibinfo{volume}{315}}, \bibinfo{pages}{69--73}, \doiprefix\url{https://doi.org/10.1016/j.optcom.2013.10.086} (\bibinfo{year}{2014}).

\bibitem{Ewoldt2015a}
\bibinfo{author}{Ewoldt, R.~H.}, \bibinfo{author}{Johnston, M.~T.} \& \bibinfo{author}{Caretta, L.~M.}
\newblock \bibinfo{title}{Experimental challenges of shear rheology: {{How}} to avoid bad data}.
\newblock In \bibinfo{editor}{Spagnolie, S.~E.} (ed.) \emph{\bibinfo{booktitle}{Complex Fluids in Biological Systems: {{Experiment}}, Theory, and Computation}}, Biological and {{Medical Physics}}, {{Biomedical Engineering}}, \bibinfo{pages}{207--241}, \doiprefix\url{https://doi.org/10.1007/978-1-4939-2065-5-6} (\bibinfo{publisher}{{Springer}}, \bibinfo{address}{{New York, NY}}, \bibinfo{year}{2015}).

\bibitem{Kerekes1992}
\bibinfo{author}{Kerekes, R.} \& \bibinfo{author}{Schell, C.}
\newblock \bibinfo{journal}{\bibinfo{title}{Characterization of {{Fibre Floccula}} tion {{Regimes}} by a {{Crowding Factor}}}}.
\newblock {\emph{\JournalTitle{Journal of Pulp and Paper Science}}} \textbf{\bibinfo{volume}{18}}, \bibinfo{pages}{32--38} (\bibinfo{year}{1992}).

\bibitem{Shafiei-Sabet2012}
\bibinfo{author}{{Shafiei-Sabet}, S.}, \bibinfo{author}{Hamad, W.~Y.} \& \bibinfo{author}{Hatzikiriakos, S.~G.}
\newblock \bibinfo{journal}{\bibinfo{title}{Rheology of {{Nanocrystalline Cellulose Aqueous Suspensions}}}}.
\newblock {\emph{\JournalTitle{Langmuir}}} \textbf{\bibinfo{volume}{28}}, \bibinfo{pages}{17124--17133}, \doiprefix\url{10.1021/la303380v} (\bibinfo{year}{2012}).

\bibitem{Decruppe1995}
\bibinfo{author}{Decruppe, J.~P.}, \bibinfo{author}{Cressely, R.}, \bibinfo{author}{Makhloufi, R.} \& \bibinfo{author}{Cappelaere, E.}
\newblock \bibinfo{journal}{\bibinfo{title}{Flow birefringence experiments showing a shear-banding structure in a {{CTAB}} solution}}.
\newblock {\emph{\JournalTitle{Colloid and Polymer Science}}} \textbf{\bibinfo{volume}{273}}, \bibinfo{pages}{346--351}, \doiprefix\url{https://doi.org/10.1007/BF00652348} (\bibinfo{year}{1995}).

\bibitem{Ito2016}
\bibinfo{author}{Ito, M.}, \bibinfo{author}{Yoshitake, Y.} \& \bibinfo{author}{Takahashi, T.}
\newblock \bibinfo{journal}{\bibinfo{title}{Shear-induced structure change in shear-banding of a wormlike micellar solution in concentric cylinder flow}}.
\newblock {\emph{\JournalTitle{Journal of Rheology}}} \textbf{\bibinfo{volume}{60}}, \bibinfo{pages}{1019--1029}, \doiprefix\url{https://doi.org/10.1122/1.4961034} (\bibinfo{year}{2016}).

\bibitem{Lane2021}
\bibinfo{author}{Lane, C.}, \bibinfo{author}{Rode, D.} \& \bibinfo{author}{R{\"o}sgen, T.}
\newblock \bibinfo{journal}{\bibinfo{title}{Two-dimensional birefringence measurement technique using a polarization camera}}.
\newblock {\emph{\JournalTitle{Applied Optics}}} \textbf{\bibinfo{volume}{60}}, \bibinfo{pages}{8435--8444}, \doiprefix\url{https://doi.org/10.1364/AO.433066} (\bibinfo{year}{2021}).

\bibitem{Koike2004}
\bibinfo{author}{Koike, Y.} \& \bibinfo{author}{Tagaya, A.}
\newblock \emph{\bibinfo{title}{Photonics polymer}} (\bibinfo{publisher}{Kyoritsu Publishing}, \bibinfo{year}{2004}).

\bibitem{Shikata1994}
\bibinfo{author}{Shikata, T.}, \bibinfo{author}{Dahman, S.~J.} \& \bibinfo{author}{Pearson, D.~S.}
\newblock \bibinfo{journal}{\bibinfo{title}{Rheo-optical behavior of wormlike micelles}}.
\newblock {\emph{\JournalTitle{Langmuir}}} \textbf{\bibinfo{volume}{10}}, \bibinfo{pages}{3470--3476}, \doiprefix\url{https://doi.org/10.1021/la00022a019} (\bibinfo{year}{1994}).

\bibitem{Santos2023}
\bibinfo{author}{Santos, T.~P.}, \bibinfo{author}{Calabrese, V.}, \bibinfo{author}{Boehm, M.~W.}, \bibinfo{author}{Baier, S.~K.} \& \bibinfo{author}{Shen, A.~Q.}
\newblock \bibinfo{journal}{\bibinfo{title}{Flow-induced alignment of protein nanofibril dispersions}}.
\newblock {\emph{\JournalTitle{Journal of Colloid and Interface Science}}} \textbf{\bibinfo{volume}{638}}, \bibinfo{pages}{487--497}, \doiprefix\url{https://doi.org/10.1016/j.jcis.2023.01.105} (\bibinfo{year}{2023}).

\bibitem{Peebles1964}
\bibinfo{author}{Peebles, F.~N.}, \bibinfo{author}{Prados, J.~W.} \& \bibinfo{author}{Honeycutt~Jr., E.~H.}
\newblock \bibinfo{journal}{\bibinfo{title}{Birefringent and rheologic properties of milling yellow suspensions}}.
\newblock {\emph{\JournalTitle{Journal of Polymer Science Part C: Polymer Symposia}}} \textbf{\bibinfo{volume}{5}}, \bibinfo{pages}{37--53}, \doiprefix\url{https://doi.org/10.1002/polc.5070050105} (\bibinfo{year}{1964}).

\bibitem{Doi1988}
\bibinfo{author}{Doi, M.}, \bibinfo{author}{Edwards, S.~F.} \& \bibinfo{author}{Edwards, S.~F.}
\newblock \emph{\bibinfo{title}{The theory of polymer dynamics}} (\bibinfo{publisher}{oxford university press}, \bibinfo{year}{1988}).

\bibitem{Maguire1980}
\bibinfo{author}{Maguire, J.~F.}, \bibinfo{author}{McTague, J.~P.} \& \bibinfo{author}{Rondelez, F.}
\newblock \bibinfo{journal}{\bibinfo{title}{Rotational diffusion of sterically interacting rodlike macromolecules}}.
\newblock {\emph{\JournalTitle{Physical Review Letters}}} \textbf{\bibinfo{volume}{45}}, \bibinfo{pages}{1891--1894}, \doiprefix\url{https://doi.org/10.1103/PhysRevLett.45.1891} (\bibinfo{year}{1980}).

\bibitem{Sdougos1984}
\bibinfo{author}{Sdougos, H.~P.}, \bibinfo{author}{Bussolari, S.~R.} \& \bibinfo{author}{Dewey, C.~F.}
\newblock \bibinfo{journal}{\bibinfo{title}{Secondary flow and turbulence in a cone-and-plate device}}.
\newblock {\emph{\JournalTitle{Journal of Fluid Mechanics}}} \textbf{\bibinfo{volume}{138}}, \bibinfo{pages}{379--404}, \doiprefix\url{https://doi.org/10.1017/S0022112084000161} (\bibinfo{year}{1984}).

\bibitem{Johnston2013}
\bibinfo{author}{Johnston, M.~T.} \& \bibinfo{author}{Ewoldt, R.~H.}
\newblock \bibinfo{journal}{\bibinfo{title}{Precision rheometry: {{Surface}} tension effects on low-torque measurements in rotational rheometers}}.
\newblock {\emph{\JournalTitle{Journal of Rheology}}} \textbf{\bibinfo{volume}{57}}, \bibinfo{pages}{1515--1532}, \doiprefix\url{https://doi.org/10.1122/1.4819914} (\bibinfo{year}{2013}).

\bibitem{Soulages2009}
\bibinfo{author}{Soulages, J.}, \bibinfo{author}{Oliveira, M. S.~N.}, \bibinfo{author}{Sousa, P.~C.}, \bibinfo{author}{Alves, M.~A.} \& \bibinfo{author}{McKinley, G.~H.}
\newblock \bibinfo{journal}{\bibinfo{title}{Investigating the stability of viscoelastic stagnation flows in {T}-shaped microchannels}}.
\newblock {\emph{\JournalTitle{Journal of Non-Newtonian Fluid Mechanics}}} \textbf{\bibinfo{volume}{163}}, \bibinfo{pages}{9--24}, \doiprefix\url{https://doi.org/10.1016/j.jnnfm.2009.06.002} (\bibinfo{year}{2009}).

\bibitem{RODD2005}
\bibinfo{author}{Rodd, L.~E.}, \bibinfo{author}{Scott, T.~P.}, \bibinfo{author}{Boger, D.~V.}, \bibinfo{author}{Cooper-White, J.~J.} \& \bibinfo{author}{McKinley, G.~H.}
\newblock \bibinfo{journal}{\bibinfo{title}{The inertio-elastic planar entry flow of low-viscosity elastic fluids in micro-fabricated geometries}}.
\newblock {\emph{\JournalTitle{Journal of Non-Newtonian Fluid Mechanics}}} \textbf{\bibinfo{volume}{129}}, \bibinfo{pages}{1--22}, \doiprefix\url{https://doi.org/10.1016/j.jnnfm.2005.04.006} (\bibinfo{year}{2005}).

\end{thebibliography}
\section*{Acknowledgements}
This work was funded by JSPS KAKENHI Grant Nos.~JP20H00222 and JP20H00223, Presto Grant No.~JPMJPR21O5, and Japan Agency for Medical Research and Development (Grant No.~JP22he0422016). We thank Prof.~T.~Rösgen, Dr.~C.~Lane, and Prof.~T.~Shikata for their valuable discussions and suggestions. We also thank Dr.~J.~Yee for providing language help and writing assistance.

\section*{Author contributions}
Y.T. and M.M. made substantial contributions to the design of the work. W.W. conducted the experiments, contributed to the data analysis and interpretation, and drafted the original manuscript. K.N. and Y.Y. made substantial contributions to the interpretation of the data and to the revision of the manuscript drafts. All authors agree to be accountable for all aspects of the work in ensuring that questions related to the accuracy or integrity of any part of the work are appropriately investigated and resolved.

\section*{Competing interests}
All authors declare that they have no known competing financial interests or personal relationships that could have appeared to influence the research reported in this paper.

\section*{Additional information}
\begin{itemize}

\item {\bf Ethics approval}:~Not applicable.

\item {\bf Availability of data and materials}:~Data sets generated during the current study are available from the corresponding author Y.T. upon reasonable request.

\end{itemize}

\end{document}